\documentclass{aastex62}

\graphicspath{{./}{Figures/}}
\submitjournal{ApJ}

\shorttitle{  Time dependence of the flux of helium nuclei with PAMELA}
\shortauthors{   N. Marcelli et al.}

\begin{document}

\title{ Time dependence of the flux of helium nuclei in cosmic rays measured by the PAMELA experiment between July 2006 and December 2009}

\correspondingauthor{R. Munini}
\email{riccardo.munini@ts.infn.it}

\author{N. Marcelli} 
\affiliation{University of Rome ``Tor Vergata'', Department of Physics, I-00133 Rome, Italy}
\affiliation{INFN, Sezione di Rome ``Tor Vergata'', I-00133 Rome, Italy} 

\author{M. Boezio}
\affiliation{INFN, Sezione di Trieste, I-34149 Trieste, Italy} 
\affiliation{IFPU, I-34014 Trieste, Italy} 

\author{A. Lenni}
\affiliation{University of Trieste, Department of Physics, I-34100 Trieste, Italy}
\affiliation{INFN, Sezione di Trieste, I-34149 Trieste, Italy} 
\affiliation{IFPU, I-34014 Trieste, Italy} 

\author{W. Menn}
\affiliation{Universitat Siegen, Department of Physics, D-57068 Siegen, Germany}

\author{R. Munini} 
\affiliation{INFN, Sezione di Trieste, I-34149 Trieste, Italy} 
\affiliation{IFPU, I-34014 Trieste, Italy}

\author{O. P. M. Aslam}, 
\affiliation{North-West University, Centre for Space Research, 2520 Potchefstroom, South Africa}

\author{D. Bisschoff} 
\affiliation{North-West University, Centre for Space Research, 2520 Potchefstroom, South Africa}

\author{M. D. Ngobeni} 
\affiliation{North-West University, Centre for Space Research, 2520 Potchefstroom, South Africa}

\author{M. S. Potgieter}
\affiliation{North-West University, Centre for Space Research, 2520 Potchefstroom, South Africa}
\affiliation{Retired from Centre for Space Research}

\author{O. Adriani} 
\affiliation{University of Florence, Department of Physics, I-50019 Sesto Fiorentino, Florence, Italy}
\affiliation{INFN, Sezione di Florence, I-50019 Sesto Fiorentino, Florence, Italy} 

\author{G. C. Barbarino} 
\affiliation{University of Naples ``Federico II'', Department of Physics, I-80126 Naples, Italy}  
\affiliation{INFN,Sezione di Naples, I-80126 Naples, Italy}

\author{G. A. Bazilevskaya} 
\affiliation{Lebedev Physical Institute, RU-119991, Moscow, Russia}

\author{R. Bellotti}
\affiliation{University of Bari, Department of Physics, I-70126 Bari, Italy}
\affiliation{INFN, Sezione di Bari, I-70126 Bari, Italy}

\author{E. A. Bogomolov}
\affiliation{Ioffe Physical Technical Institute, RU-194021 St. Petersburg, Russia} 

\author{M. Bongi} 
\affiliation{University of Florence, Department of Physics, I-50019 Sesto Fiorentino, Florence, Italy}
\affiliation{INFN, Sezione di Florence, I-50019 Sesto Fiorentino, Florence, Italy}

\author{V. Bonvicini} 
\affiliation{INFN, Sezione di Trieste, I-34149 Trieste, Italy}

\author{A. Bruno} 
\affiliation{Heliophysics Division, NASA Goddard Space Flight Center, Greenbelt, MD, USA}
\affiliation{Department of Physics, Catholic University of America, Washington DC, USA}

\author{F. Cafagna} 
\affiliation{INFN, Sezione di Bari, I-70126 Bari, Italy}

\author{D. Campana} 
\affiliation{INFN,Sezione di Naples, I-80126 Naples, Italy} 

\author{P. Carlson} 
\affiliation{KTH, Department of Physics, and the Oskar Klein Centre for Cosmoparticle Physics,AlbaNova University Centre, SE-10691 Stockholm, Sweden}

\author{M. Casolino}
\affiliation{INFN, Sezione di Rome ``Tor Vergata'', I-00133 Rome, Italy}  
\affiliation{RIKEN, EUSO team Global Research Cluster, Wako-shi, Saitama, Japan} 

\author{G. Castellini} 
\affiliation{IFAC, I-50019 Sesto Fiorentino, Florence, Italy}

\author{C. De Santis}
\affiliation{INFN, Sezione di Rome ``Tor Vergata'', I-00133 Rome, Italy}  

\author{A. M. Galper}
\affiliation{MEPhI: National Research Nuclear University MEPhI, RU-115409, Moscow, Russia} 

\author{S. V. Koldashov} 
\affiliation{MEPhI: National Research Nuclear University MEPhI, RU-115409, Moscow, Russia} 

\author{S. Koldobskiy} 
\affiliation{MEPhI: National Research Nuclear University MEPhI, RU-115409, Moscow, Russia}

\author{A. N. Kvashnin}
 \affiliation{Lebedev Physical Institute, RU-119991, Moscow, Russia}

\author{A.A. Leonov} 
\affiliation{MEPhI: National Research Nuclear University MEPhI, RU-115409, Moscow, Russia} 

\author{V.V. Malakhov} 
\affiliation{MEPhI: National Research Nuclear University MEPhI, RU-115409, Moscow, Russia} 

\author{L. Marcelli} 
\affiliation{INFN, Sezione di Rome ``Tor Vergata'', I-00133 Rome, Italy}  

\author{M. Martucci} 
\affiliation{University of Rome ``Tor Vergata'', Department of Physics, I-00133 Rome, Italy}
\affiliation{INFN, Laboratori Nazionali di Frascati, Via Enrico Fermi 40, I-00044 Frascati, Italy}

\author{A. G. Mayorov} 
\affiliation{MEPhI: National Research Nuclear University MEPhI, RU-115409, Moscow, Russia}

\author{M. Merg\`e} 
\affiliation{INFN, Sezione di Rome ``Tor Vergata'', I-00133 Rome, Italy}  
\affiliation{University of Rome ``Tor Vergata'', Department of Physics, I-00133 Rome, Italy}

\author{E. Mocchiutti} 
\affiliation{INFN, Sezione di Trieste, I-34149 Trieste, Italy} 

\author{A. Monaco} 
\affiliation{University of Bari, Department of Physics, I-70126 Bari, Italy}
\affiliation{INFN, Sezione di Bari, I-70126 Bari, Italy}

\author{N. Mori} 
\affiliation{INFN, Sezione di Florence, I-50019 Sesto Fiorentino, Florence, Italy} 

\author{V. V. Mikhailov} 
\affiliation{MEPhI: National Research Nuclear University MEPhI, RU-115409, Moscow, Russia}

\author{G. Osteria}
\affiliation{INFN,Sezione di Naples, I-80126 Naples, Italy}

\author{B. Panico} 
\affiliation{INFN,Sezione di Naples, I-80126 Naples, Italy}

\author{P. Papini} 
\affiliation{INFN, Sezione di Florence, I-50019 Sesto Fiorentino, Florence, Italy} 

\author{M. Pearce}
\affiliation{KTH, Department of Physics, and the Oskar Klein Centre for Cosmoparticle Physics,AlbaNova University Centre, SE-10691 Stockholm, Sweden}

\author{P. Picozza} 
\affiliation{INFN, Sezione di Rome ``Tor Vergata'', I-00133 Rome, Italy}  
\affiliation{University of Rome ``Tor Vergata'', Department of Physics, I-00133 Rome, Italy}

\author{M. Ricci}
\affiliation{INFN, Laboratori Nazionali di Frascati, Via Enrico Fermi 40, I-00044 Frascati, Italy}

\author{S. B. Ricciarini}
\affiliation{INFN, Sezione di Florence, I-50019 Sesto Fiorentino, Florence, Italy} 
\affiliation{IFAC, I-50019 Sesto Fiorentino, Florence, Italy}

\author{M. Simon}
\affiliation{Universitat Siegen, Department of Physics, D-57068 Siegen, Germany}
\affiliation{Deceased}

\author{A. Sotgiu}
\affiliation{INFN, Sezione di Rome ``Tor Vergata'', I-00133 Rome, Italy}

\author{R. Sparvoli}
\affiliation{INFN, Sezione di Rome ``Tor Vergata'', I-00133 Rome, Italy} 
\affiliation{University of Rome ``Tor Vergata'', Department of Physics, I-00133 Rome, Italy}

\author{P. Spillantini}
\affiliation{MEPhI: National Research Nuclear University MEPhI, RU-115409, Moscow, Russia} 
\affiliation{Istituto Nazionale di Astrofisica, Fosso del cavaliere 100, 00133 Roma, Italy} 

\author{Y. I. Stozhkov} 
 \affiliation{Lebedev Physical Institute, RU-119991, Moscow, Russia} 

\author{A. Vacchi}
\affiliation{INFN, Sezione di Trieste, I-34149 Trieste, Italy} 
\affiliation{University of Udine, Department of Mathematics, Computer Science and Physics Via delle Scienze, 206, Udine, Italy}

\author{E. Vannuccini}
\affiliation{INFN, Sezione di Florence, I-50019 Sesto Fiorentino, Florence, Italy} 

\author{G.I. Vasilyev} 
\affiliation{Ioffe Physical Technical Institute, RU-194021 St. Petersburg, Russia} 

\author{S. A. Voronov} 
\affiliation{MEPhI: National Research Nuclear University MEPhI, RU-115409, Moscow, Russia} 

\author{Y. T. Yurkin} 
\affiliation{MEPhI: National Research Nuclear University MEPhI, RU-115409, Moscow, Russia} 

\author{G. Zampa}
\affiliation{INFN, Sezione di Trieste, I-34149 Trieste, Italy} 

\author{N. Zampa}
\affiliation{INFN, Sezione di Trieste, I-34149 Trieste, Italy}

\begin{abstract}

Precise time-dependent measurements of the Z = 2 component in  the cosmic radiation provide crucial information about  the propagation of charged particles
through the heliosphere. The PAMELA experiment, with its long flight duration (15$^{\rm th}$ June 2006 - 23$^{\rm rd}$ January 2016) and the low energy threshold 
($80$ MeV/n)  is an ideal detector for cosmic ray solar modulation studies. In this paper, the helium nuclei spectra measured by the PAMELA  instrument 
from July 2006 to December 2009 over a Carrington rotation time basis are presented. A state-of-the-art three-dimensional model for  cosmic-ray propagation 
inside the heliosphere was used to interpret the time-dependent measured fluxes. 
Proton-to-helium flux ratio time profiles at various rigidities are also presented in order to study any features which could result from the different 
masses and local interstellar spectra shapes.

\end{abstract}


\section{Introduction} \label{sec:intro}

Helium nuclei are the most abundant component of galactic cosmic rays (CRs) besides protons. They represent approximately $9\%$ of the total CR budget and together with 
protons account for $98\%$ of the cosmic radiation. The vast majority of helium nuclei are 
believed to be accelerated at astrophysical sources like supernovae
remnants. 
Then, for several million of years before reaching the Earth's Solar
System, 
CRs had propagated 
through the Galaxy interacting with the interstellar matter  
and diffusing on the 
galactic magnetic field. 
These processes modify the CR spectral shapes and compositions with respect to the acceleration site. 
The precise measurement of the CR helium nuclei over an
extended energy 
range  
is crucial in order to better understand both their origin and
propagation history (e.g. \cite{ama18}). The most recent and
most precise 
measurements of CR helium and  
proton spectra, provided by experiments such as PAMELA
\citep{Adriani69}, AMS-02
\citep{PhysRevLett.114.171103,PhysRevLett.115.211101}, CREAM
\citep{Yoon_2017} and ATIC \citep{ATIC}, highlighted 
features in these spectra that challenged the commonly accepted scenario
of CR acceleration and propagation in the Galaxy. 
However, the vast majority of CR measurements were obtained well
inside the heliosphere where solar modulation affects the CR spectra.  

When entering the heliosphere, CRs encounter the turbulent heliospheric
magnetic field (HMF) embedded into the solar wind. Particles are scattered by HMF irregularities
and undergo convection, diffusion, adiabatic deceleration and drift
motions due to the presence of HMF gradients and curvatures. 
As a consequence, mostly below a few tens of GV, the CR intensities and
spectral shapes change with respect to the Local Interstellar Spectra
(LIS), i.e. the spectra which would be measured outside the  
heliospheric boundaries. This effect is known as solar modulation (e.g. \cite{Potgieter_2013,pot17}).
The solar activity follows an $11$-year cycle which introduces a time
dependence:  
higher CR intensity is measured during solar minimum, lower during
solar maximum.  
Precise time-dependent measurements of low energy CRs improve the
understanding of solar modulation. These studies  
are essential to calibrate models that describe mechanisms for the modulation of CRs in the heliosphere (e.g. \cite{pot13}). Furthermore, sophisticated   
models of particle transport in the heliosphere play a fundamental
role in space weather 
and to predict radiation
hazard for long-term spacecraft missions. 
 
The PAMELA detector, due to its low energy limit and long flight duration, represents an ideal instrument
to perform precise measurements of the time-dependent fluxes of different CR species. 
The PAMELA collaboration already published several 
papers on CR solar modulation: protons \citep{Adriani_2013,Martucci_2018}, electrons \citep{Adriani_el_2015}
and positron-to-electron flux ratio \citep{Adriani_elpos_2016}. In this
paper the time-dependent helium spectra measured by the PAMELA
instrument during the 
$23^{\rm rd}$ solar minimum (July 2006 - December 2009) 
are presented. The
extraordinary quiet solar modulation period from 2006 to 2009 (e.g. \cite{pot15}) provides excellent opportunity to study CR
propagation in the heliosphere under relative perfect modulation
conditions (e.g. \cite{difelice}). 
The spectra were measured for each Carrington rotation period 
($\approx 27$ days)
assuring an optimal
compromise between a good time resolution and a high statistic. The studied
time  period  
 corresponds to Carrington rotation numbers $2045-2091$. No
isotopic separation 
was performed in this work, consequently the fluxes correspond to the  
sum of $^4$He and $^3$He fluxes. 

Additionally, helium nuclei fluxes are compared to that for protons, as a function of time and rigidity.   
From a
solar modulation point of view, time-dependent proton-to-helium flux
ratios
may result from the different particle
masses and from the different shapes 
of the local interstellar spectra \citep{tom18,Cor19}. 

Finally, the modelling of these reported spectra, with a state-of-the-art 3D numerical model is presented.
Excellent qualitative and good quantitative agreement between modelling results and the observed spectra is found.
 
\section{The PAMELA detector} \label{sec:pamela}

\begin{figure}
\centering
\includegraphics[width=.5\textwidth]{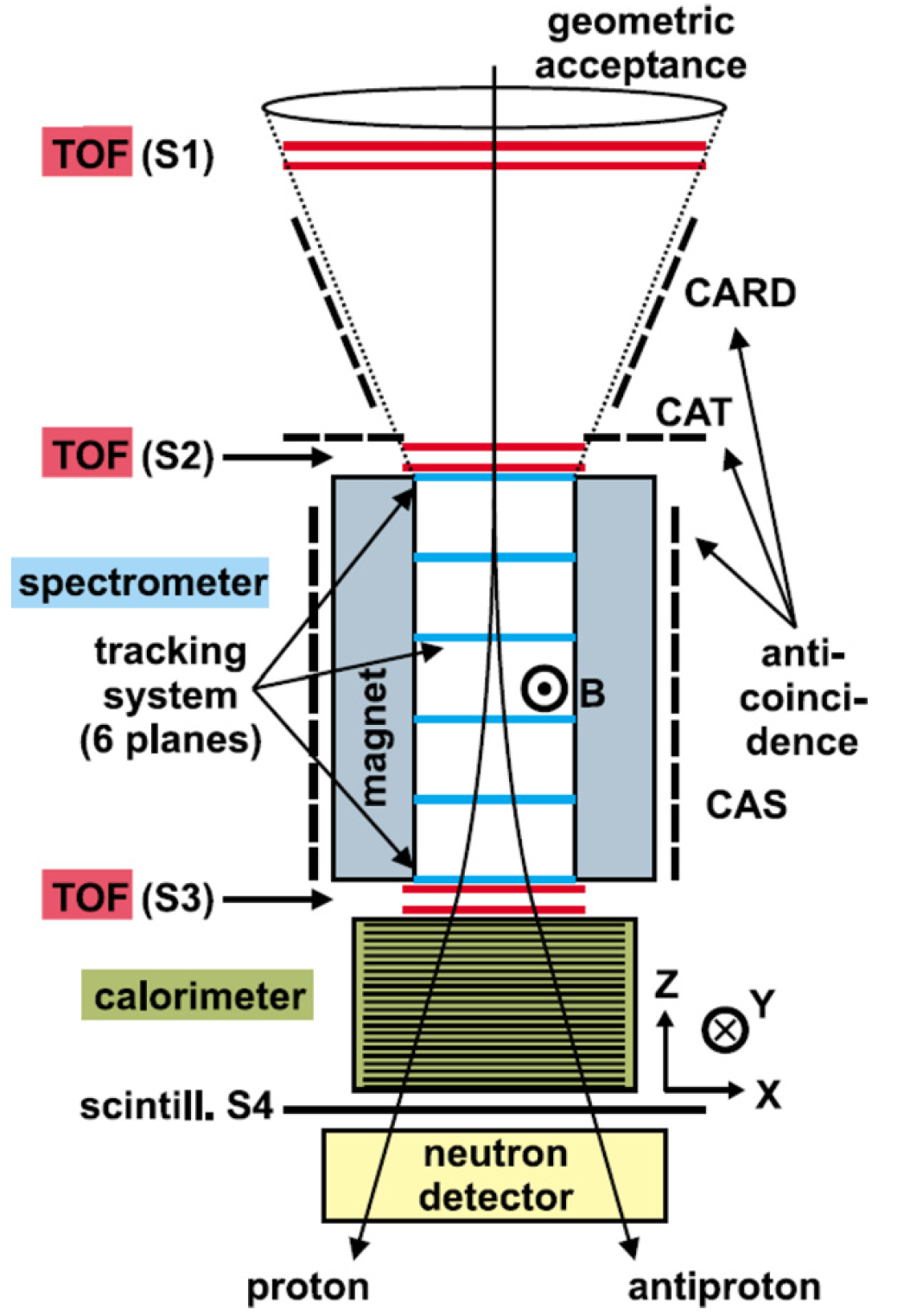}
\caption{PAMELA and its sub-detectors.}
\label{fig_PAM}
\end{figure}

PAMELA (Payload for Antimatter-Matter Exploration and Light-nuclei Astrophysics) is a satellite-borne experiment designed to make long duration observations of the cosmic
radiation from Low Earth Orbit \citep{Picozza_2007}. The PAMELA payload is schematically shown in Figure \ref{fig_PAM}.
 The instrument collected galactic CRs for almost 10 years from 2006 June 15 when it was launched from the Baikonur cosmodrome
in Kazakhstan, up to January 2016. The PAMELA instrument is hosted on board of the Russian satellite Resurs-DK1 with an elliptical orbit at an altitude ranging between 350 and 
610 km with an inclination of $70^{\circ}$. After 2010 the orbit was changed to a circular one at a constant altitude of about $600$ km.

The payload comprises a number of redundant detectors capable of identifying particles by providing charge, rigidity and velocity measurements over a wide energy 
range. Multiple sub-detectors are built around a magnetic spectrometer, composed of a silicon tracking system \citep{Adriani_2003} placed inside a $0.43\ $T permanent magnet.
The $300$ $\mu$m thick double-sided Silicon sensors of the tracking system measure two impact coordinates on each plane, reconstructing with high accuracy the particle
deflection with a maximum detectable rigidity of $\sim1.2$ TV, and the absolute electric charge up to  $Z \approx 8$. 

A time of flight system (ToF) is composed of six layers of plastic scintillators, arranged in three double planes (S1, S2, and S3).
The ToF,  with a resolution of $\sim300$ ps, provides a fast signal for data acquisition \citep{Osteria_2004}. The trigger configuration is defined by coincidental energy deposit in S1, S2 and S3. 
The ToF also contributes to
particle identification measuring through the ionization energy loss the absolute charge up to $Z \approx 6$  and the particle 
velocity and incoming direction that, together with the curvature measured by the tracking system, allows the charge sign to be determined.

An electromagnetic imaging W/Si calorimeter ($16.3$ radiation lengths and $0.6$ interaction lengths deep) provides hadron-lepton discrimination \citep{Boezio_2002}. A neutron
counter \citep{Stozhkov_2005} contributes to discrimination power by detecting the increased neutron production in the calorimeter associated with hadronic showers compared
to electromagnetic
ones, while a plastic scintillator, placed beneath the calorimeter, increases the identification of high-energy electrons. The whole apparatus is surrounded by an
anti-coincidence system
(AC) of three sets of scintillators (CARD, CAS, and CAT) for the rejection of background events \citep{Orsi_2005}. A comprehensive description of the instrument can be found
in \cite{Adriani_PhysRep_2014}.

\section{Analysis}

\subsection{Event Selection}
\begin{figure}
\centering
\includegraphics[width=1\textwidth]{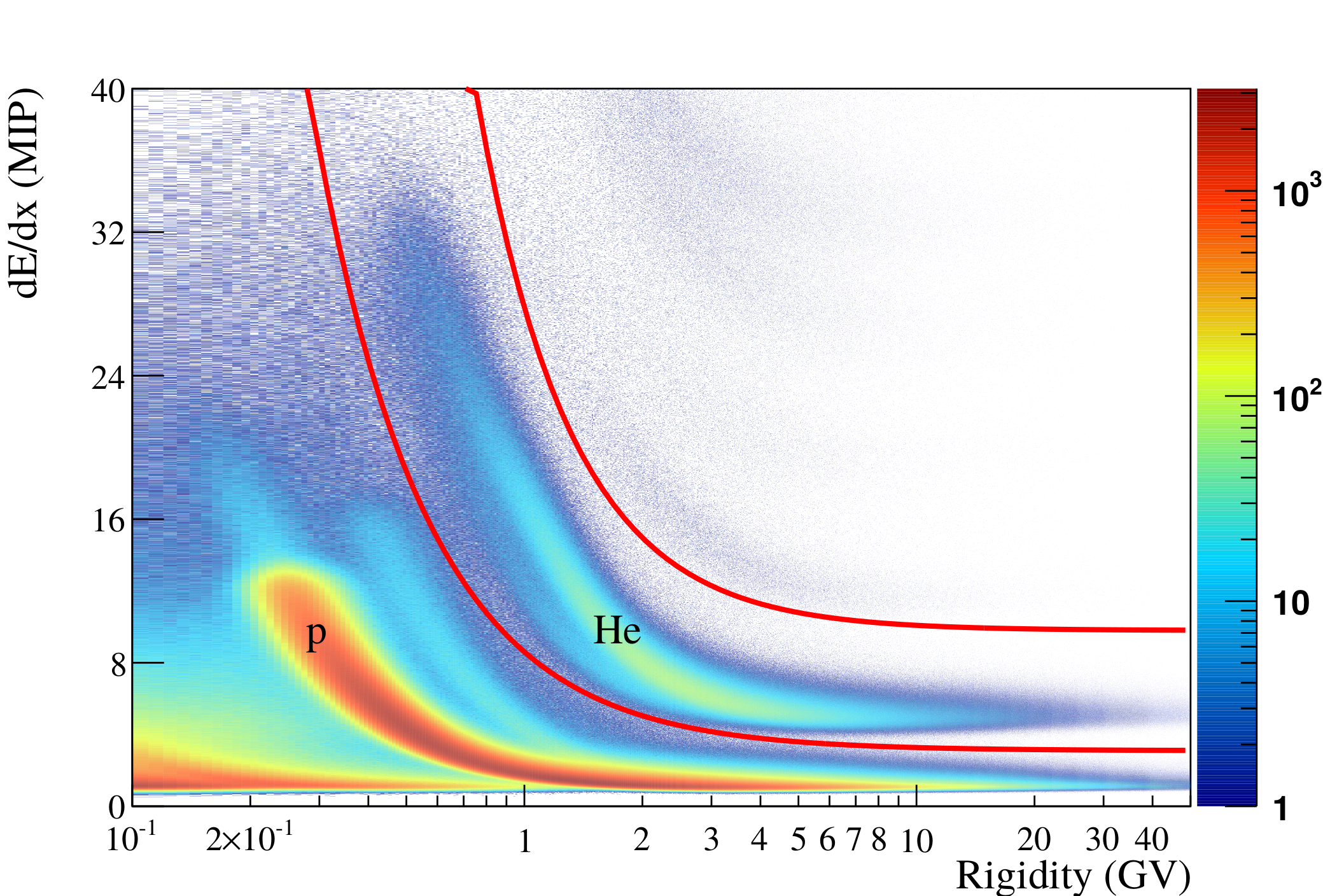}
\caption{Distribution of the  average
ionization energy loss in terms of the minimum ionizing
particle (MIP) inside the silicon tracker planes as a function of
particle rigidity. The helium nuclei band is clearly separated from
the hydrogen nuclei one.}
\label{figdistr}
\end{figure}

A clean sample of helium nuclei events was obtained applying cuts on the information provided by the PAMELA sub-detectors:
\begin{enumerate}

\item Only events with a single track fitted in the spectrometer and fully contained within $1.5 $ mm away from the magnet walls 
and the ToF-scintillators edges were selected. The track of the selected events had to have a lever-arm of at least 4 silicon planes
and a minimum of 3 hits on both the bending \textit{x}-view and
non-bending \textit{y}-view.
Low-quality tracks were rejected requiring a good $\chi^2$ resulting from the fit.

\item Down-going particles were selected requiring a positive $\beta=\frac{v}{c}$ ($v$ particle speed, $c$ speed of light) measured by the ToF system.
This selection rejected splash albedo particles.


\item Helium nuclei were finally selected according to their mean energy release  in terms of the minimum ionizing
particle (MIP)\footnote{Energy loss is expressed in terms of MIP, which is the energy released by a
particle for which the mean energy loss rate in matter is minimum.} over the silicon planes of the tracking system. Only events which were consistent with the helium band 
 shown in Figure \ref{figdistr} were selected. Because of the energy
losses, the minimum detectable rigidity for helium was on average
$840$ MV, i.e.  $ \sim 90$ MeV/n    in kinetic energy. Below this energy,  
helium nuclei were not able to reach S3 and trigger the data acquisition.

\end{enumerate}

No attempt was made in this paper to separate between $^3$He and $^4$He nuclei, although separation is possible between 0.1 and 1.5 GeV/n  
\citep{Adriani_2016}. Hence, in the analysis all Z=2 particles were treated as being $^4$He.

\subsection{\textit{Selection efficiencies}}


The redundant information provided by the PAMELA sub-detectors allowed to study the selection efficiencies with flight data. 
To account for any possible time variation of the detector response, the efficiencies were evaluated for each Carrington rotation. 
Simulated data were used to cross check the flight efficiencies.
With the Monte Carlo data it was possible to reproduce and study
all selection efficiencies, their rigidity and time dependence allowing to detect
possible sources of bias, like contamination
of efficiency samples and correlation among selection criteria.

The tracking system efficiency (criterion 1) and its energy
dependence were obtained by Monte Carlo data. This efficiency was found to decrease over the years from a maximum of $\approx 90\%$ in $2006$ to $\approx 20\%$ at the end
of $2009$. This significant time dependence
was due to the sudden, random failure of a few front-end chips in the tracking system.
This resulted in a progressive reduction of the number of hits
available for track reconstruction which in turn produced a decreasing efficiency. The front-end chip failure was properly simulated  
with the inclusion in the Monte Carlo of a time-dependent map of dead channels. In order to decrease the statistical fluctuations, each efficiency was fitted with an 
appropriate function that reproduced its  rigidity dependence. The values resulting from the fit were used to calculate the fluxes. Two samples of $^4$He and $^3$He were simulated in order 
 to account for possible differences in the tracking efficiency for the two isotopes but no significant differences were found. 
 
Similarly to the procedure adopted for the time-dependent analysis of the cosmic-ray protons \citep{Adriani_2013,Martucci_2018},
in order to account for any residual instrumental time-dependence,  
the fluxes of each Carrington rotation were normalized between $\approx 20$
and $50$ GV\footnote{This amounts to assume that solar modulation effects are
negligible in this rigidity range.}
to the flux measured in the same rigidity region 
over the period July 2006 - March 2008 \citep{Adriani_2011}. 

Flight efficiency for criterion $3$ was obtained from a sample of
helium nuclei selected by means of the ionization energy loss
information provided by the ToF system.
The selection efficiency was found to be
constant with time and higher than $99\%$ in the whole rigidity range.


\subsection{Statistical deconvolution }

The helium rigidity measured with the tracking system differs from the initial rigidity at the top of the payload. 
This is due to the finite spectrometer resolution and the ionization energy losses suffered by the particles traversing the instrument. 
This latter effect was particularly relevant for helium nuclei below 2 GeV. 
A correction was applied by means of an unfolding procedure, following the Bayesian
approach proposed by \cite{DAgostini_1995}. A detector response matrix was calculated with Monte Carlo data, one
for each Carrington rotation under analysis. 
The unfolding procedure was applied to the count distributions of selected events
binned according to their measured rigidities and divided by all selection efficiencies except
those of the tracking system. The tracking system efficiency was instead applied to the unfolded count distribution.

\subsection{Flux determination}

The absolute helium flux $\Phi(E)$ in kinetic energy was obtained as follows: 
\begin{equation}
    \Phi(E)=\frac{N(E)}{G(E) \times T \times \epsilon(E)\times \Delta E}
\end{equation}

where $N(E)$ is the unfolded count distribution, $\epsilon(E)$ the efficiency of the tracking system selections,
$G(E)$ the geometrical factor, $T$ the live time 
and $\Delta E$ the width of the energy interval.  For the conversion from rigidity to kinetic
energy all helium nuclei events were treated as $^4$He.

Because of the wide geomagnetic region spanned by the satellite over its orbit, the helium nuclei spectrum was evaluated for various, seven, 
vertical geomagnetic cutoff intervals, estimated using the satellite position and the St\"{o}rmer approximation. For each spectrum, only the 
fluxes estimated in energy regions at least 1.3 times above the maximum vertical geomagnetic cutoff of each interval were assumed to be of galactic
origin. Finally, the helium nuclei spectrum was determined by combining such fluxes of each geomagnetic cutoff interval weighted for the 
corresponding fractional live time.

\begin{figure}[t]
\centering
\includegraphics[width=1.\textwidth]{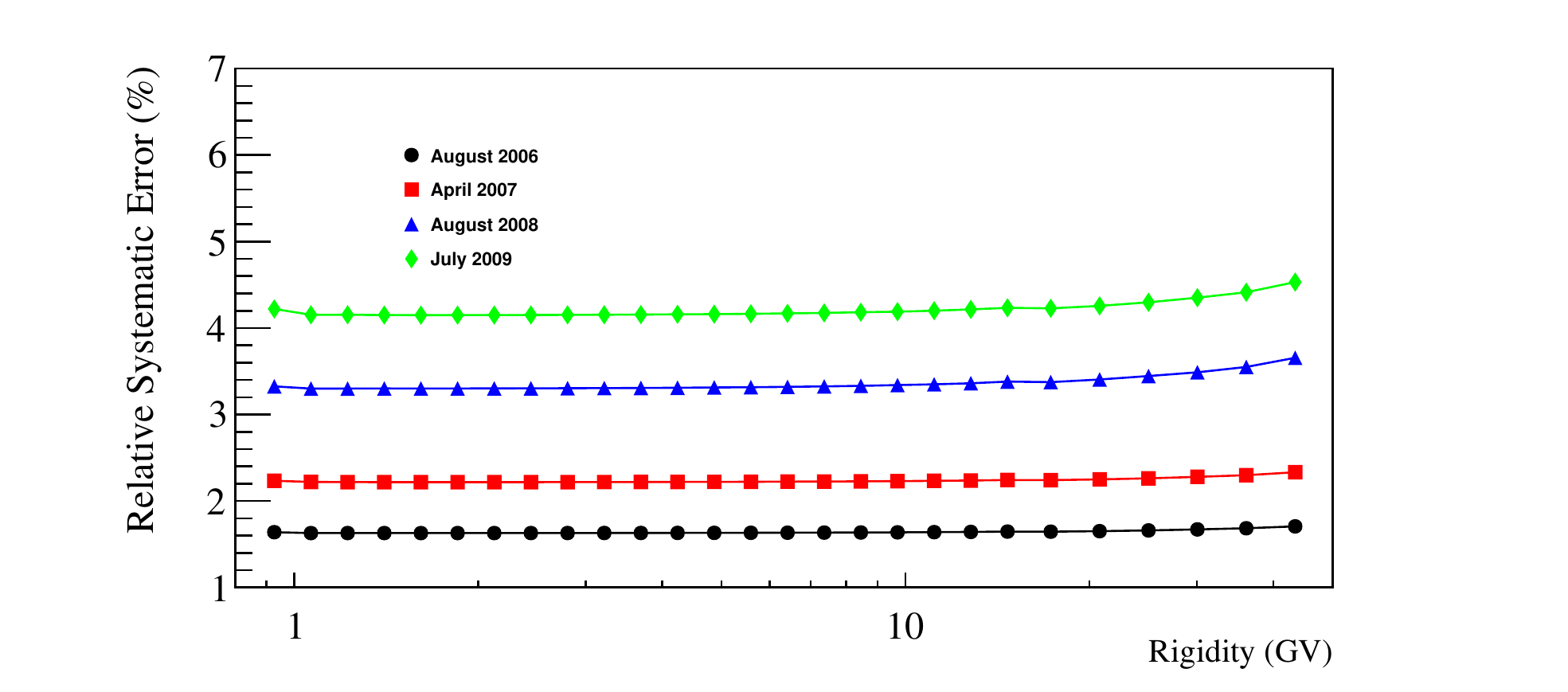}
\caption{Relative systematic error as a function of rigidity for four time intervals.}
\label{fig_syst}
\end{figure}

\subsubsection{Live time}
\label{sec:livetime}

The live time was provided by an on-board clock that timed the periods during which
the apparatus was waiting for a trigger.
The acquisition time was evaluated for each energy bin as 
the total live time spent above the geomagnetic cutoff and outside the South Atlantic Anomaly.
The accuracy of the live time determination was cross-checked by
comparing different clocks available in flight, which showed a relative difference of less than $0.2\%$.
The total live time was about $5\times10^7$ s above $∼ 20$ GV, reducing
to about $18\%$ of this value at $800$ MV because of the relatively short time spent by the
satellite at high geomagnetic latitudes.

\subsubsection{Geometrical factor}

The geometrical factor for this analysis was determined by the requirement of triggering and containment within the fiducial volume (criterion 1). 
The pure geometrical acceptance above $2$ GV is $19.5$ cm$^2$ sr. The geometrical factor, including interactions and energy losses, was estimated with the 
full simulation of the apparatus, and was found to be about $13\%$ lower because of fragmentation of helium nuclei in the instrument material above S3. 
Below 1 GV the acceptance sharply decreased because of the fluctuations in energy losses due to the particle slowdown.

\begin{figure}[t]
\centering
\includegraphics[width=1.\textwidth]{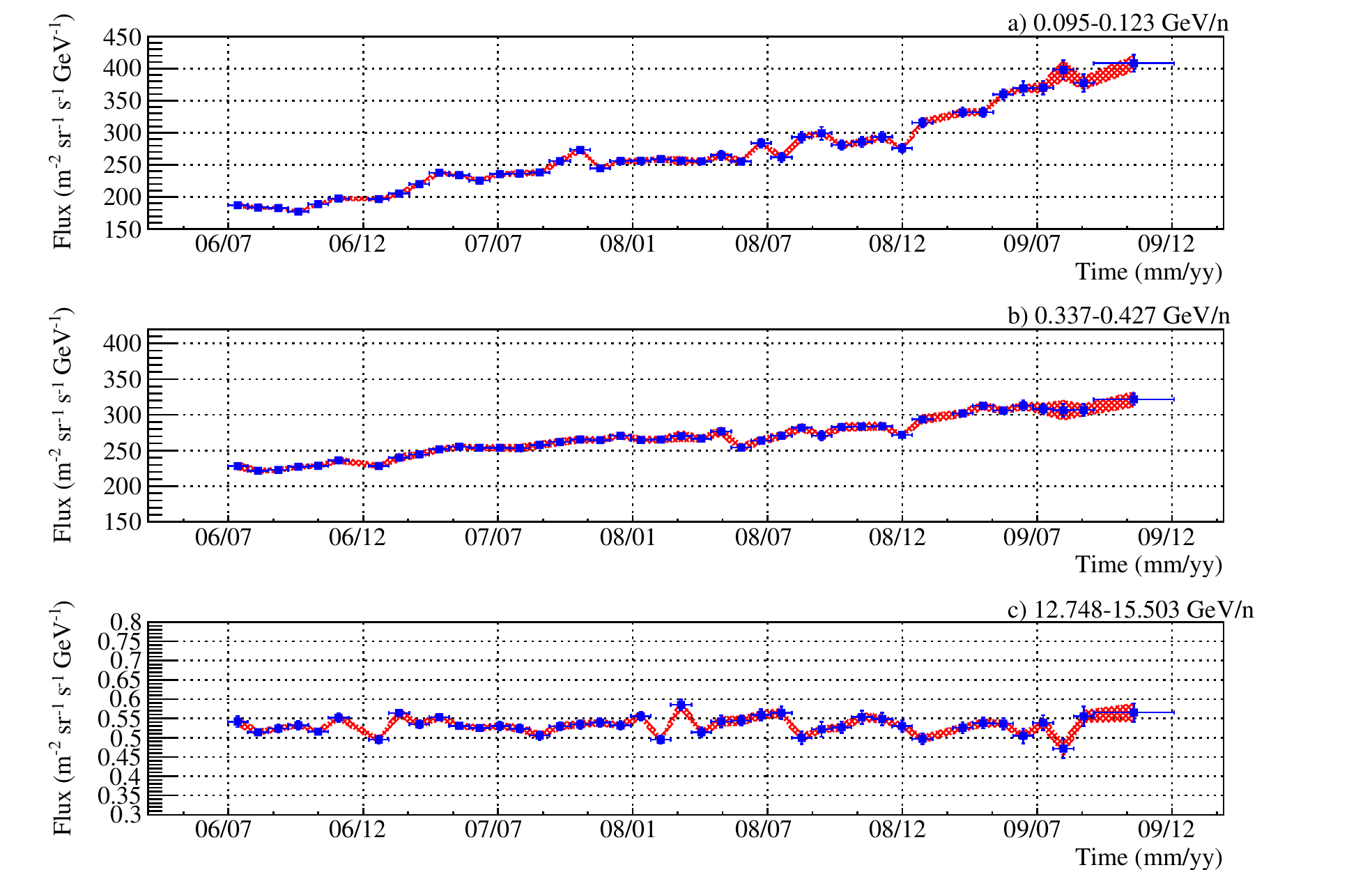}
\caption{Time profiles of the helium flux for three kinetic energy intervals; the error bars are statistical, 
while the shaded area represents the quadratic sum of all systematic uncertainties.}
\label{fig_time}
\end{figure}

\section{Systematic uncertainties}
\label{s:syst}

Various sources of systematic uncertainties were analyzed: 
\begin{enumerate}
 \item The statistical errors resulting from the finite size of the efficiency samples. Most of the efficiencies were fitted in order to reduce
 the statistical fluctuations. For each efficiency a systematic
uncertainty related to the fit estimation was obtained evaluating 
the one sigma ($68\%$) confidence intervals
associated
to the fitted values.  
 \item The live time uncertainty. As discussed in Section \ref{sec:livetime} the systematic uncertainty related to the live time was 
 estimated to be constant at $0.2\%$ \citep{Adriani_2013}.
 \item The systematic error due to the unfolding procedure. This was evaluated by folding and
unfolding a known spectral shape. A large sample of helium nuclei was simulated with an input
spectrum consistent with the reconstructed experimental spectrum at the top of the payload.
The rigidities of the simulated events were reconstructed
and the count distribution was unfolded
and compared with the initial simulated sample by means of pull distributions.
More details on this procedure are reported by \cite{Adriani_el_2015}. A systematic uncertainty 
of $\approx 2 \%$ was estimated.
 \item The error associated to the flux normalization factor. This was
assumed to be the statistical error
 on the ratio between the measured
flux ($\approx 20-50$ GV) and the corresponding flux presented by \cite{Adriani_2011}.
\end{enumerate}
The overall systematic uncertainty associated to the fluxes was obtained as the quadratic sum of all the contributions.
Figure \ref{fig_syst} shows the time dependence of the global systematic uncertainty as a function of the rigidity. The 
uncertainty increase over time was mainly due to the decreasing
efficiency of the tracking system. 

\begin{figure}
\centering
\includegraphics[width=.98\textwidth]{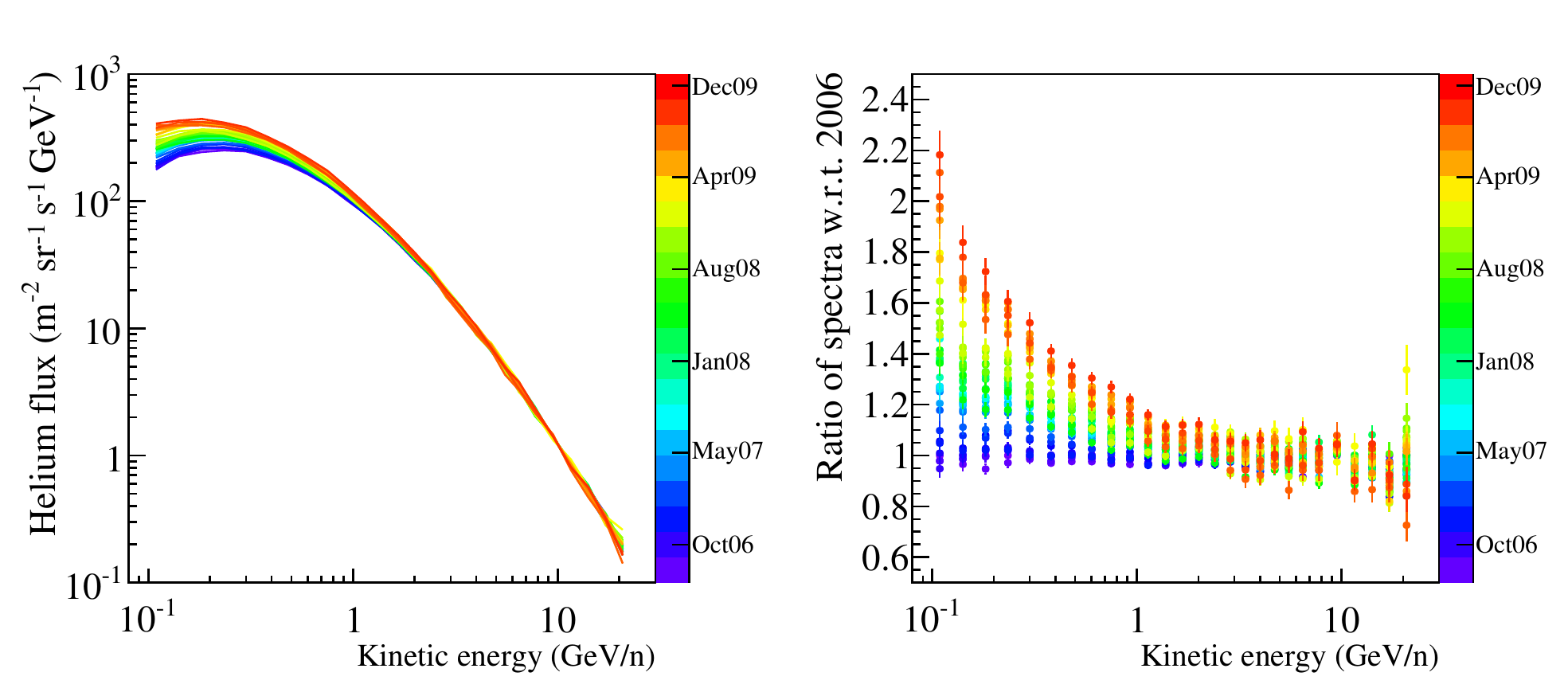}
\caption{Left panel: the evolution of the helium energy spectrum as particle intensities approached the period of minimum solar activity, from July 2006 (violet), 
to December 2009 (red). Right panel: the ratio of the measured spectra
with respect to the spectrum of July 2006. Different time periods are
color-coded on the right of each panel.  
The error bars are the result of the quadratic sum of statistical and systematic errors. }
\label{fig_rainbow_1}
\end{figure}

\section{Results}

Figure \ref{fig_time} shows the time profile of the helium flux for three kinetic energy intervals. The last four Carrington rotations of 2009 
were combined in a single spectrum for statistical reasons.  A total of 42 spectra were measured.  The error bars are statistical, while the 
shaded areas represent the quadratic sum of all systematic uncertainties. These plots show the different amount of solar modulation as a function of the energy.
The flux measured at about 100 MeV/n increased by a factor 2 from middle $2006$ to the end of $2009$
while the flux at $600$ MeV/n increased only by a factor $1.25$. Above $\approx 15$ GeV/n   
the effect of modulation was lower than the statistical precision of the measurement. 

Figure \ref{fig_rainbow_1}, left panel, shows the time evolution of helium energy spectra from July 2006 (violet curve) to December 2009 (red curve).
 Figure \ref{fig_rainbow_1}, right panel, shows the ratio
of the helium fluxes with respect to the fluxes measured in July $2006$. 
Besides the time-dependent intensity, a varying spectral shape was also observed over time. 
In $2006$ the maximum flux intensity was reached at about $250$ MeV/n slowly 
decreasing to about $150$ MeV/n by the end of $2009$. Table \ref{tab1} presents the helium nuclei spectra measured over four time
periods (Carrington rotations 2050, 2064, 2077 and 2088-2091). 

Finally the solar modulation of the helium nuclei was compared to the
published PAMELA proton spectrum evaluated during the same time period
\citep{Adriani_2013}. Also for the  
protons the last four Carrington rotations of 2009 were combined to
form an average spectrum. 

\begin{figure}
\centering
\includegraphics[width=.95\textwidth]{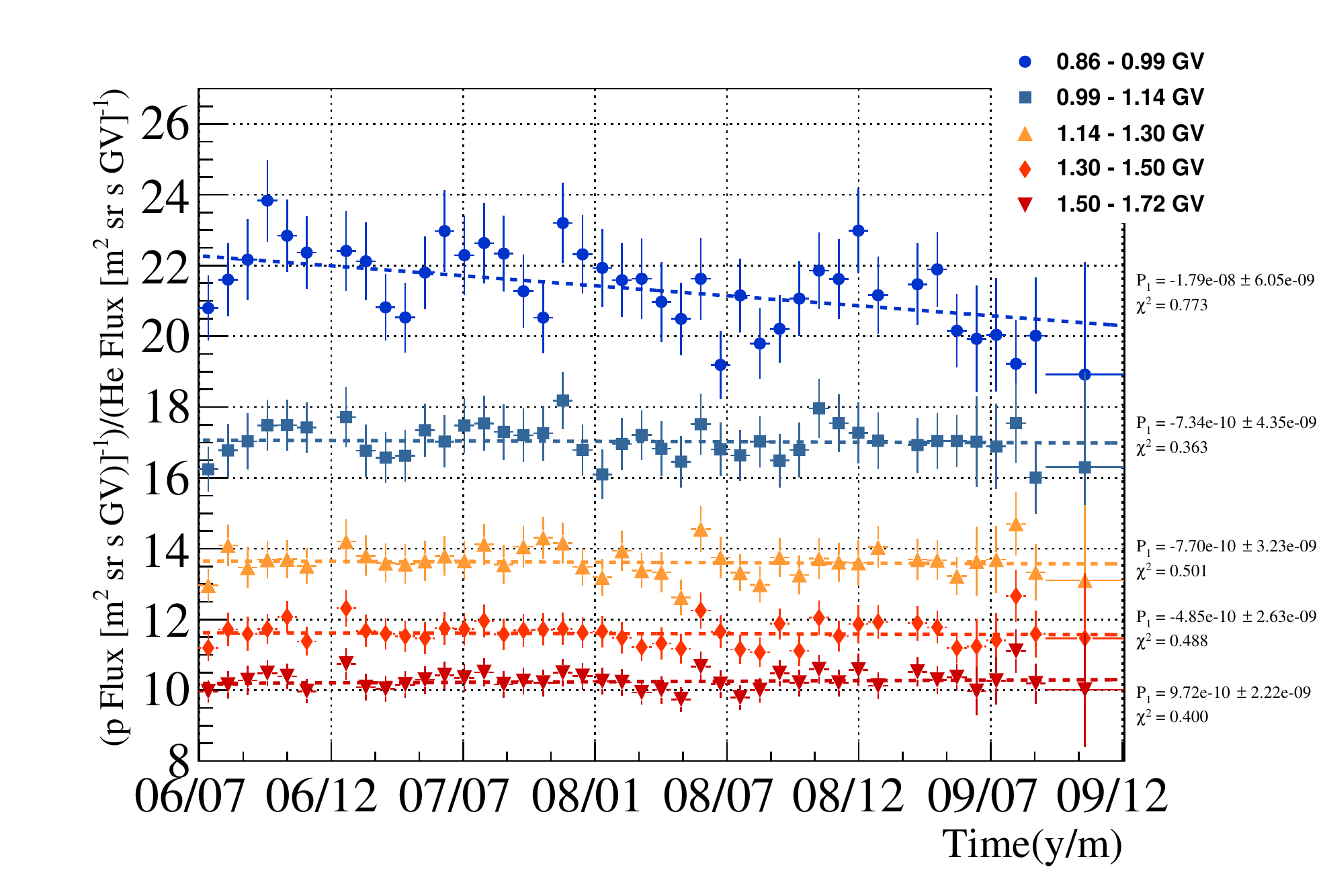}
\caption{Time profiles of proton-to-helium flux ratio in five
different rigidity intervals reported in the legend.
The error bars are the result of the quadratic sum of statistical and systematic errors. For more details see the text. }
\label{fig_phe_ratio}
\end{figure}

Figure \ref{fig_phe_ratio} shows the proton-to-helium flux ratio as a function of time for five different rigidity intervals. 
The proton and helium fluxes
were estimated as a function of rigidity, i.e. particles per m$^2$ sr s
GV.
Since the quantity measured by the magnetic spectrometer is rigidity,
this approach allows a
more precise estimation of the ratios considering that all systematic
uncertainties, related to the same instrumental effects, cancel
out. The residual systematic uncertainty includes only the errors due
to the efficiency estimation. The error bars in Figure  \ref{fig_phe_ratio} represent the
quadratic sum of the statistical errors with this residual systematic
error.

A non constant ratio points to modulation differences between protons and helium  which is further discussed below.  
The colored lines on Figure \ref{fig_phe_ratio} represent a fit with a first 
degree polynomial function of the form $p_0 + t \cdot p_1$ where $p_0$ is a normalization value, $p_1$ quantifies the amount of time dependence and $t$ is time.
Above $1$ GV the proton-to-helium flux ratio over time was well described by a constant value. On the 
contrary, below $1$ GV the fit indicates an overall decrease of
about $10\%$ from 2006 to 2009. 
However,  
because of the relatively large errors below $1$ GV,
a time independence of the proton-to-helium flux ratio cannot be excluded
also at these low rigidities. 

In addition to the long-term decrease, below 1 GV the proton-to-helium flux
ratio shows a cyclic variation with maxima registered during October 2006,
August 2007, December 2007, August 2008 and March 2009.  These short-term variations are beyond the scope of this paper and will be addressed in a future work.

\begin{figure}
\centering
\includegraphics[width=.95\textwidth]{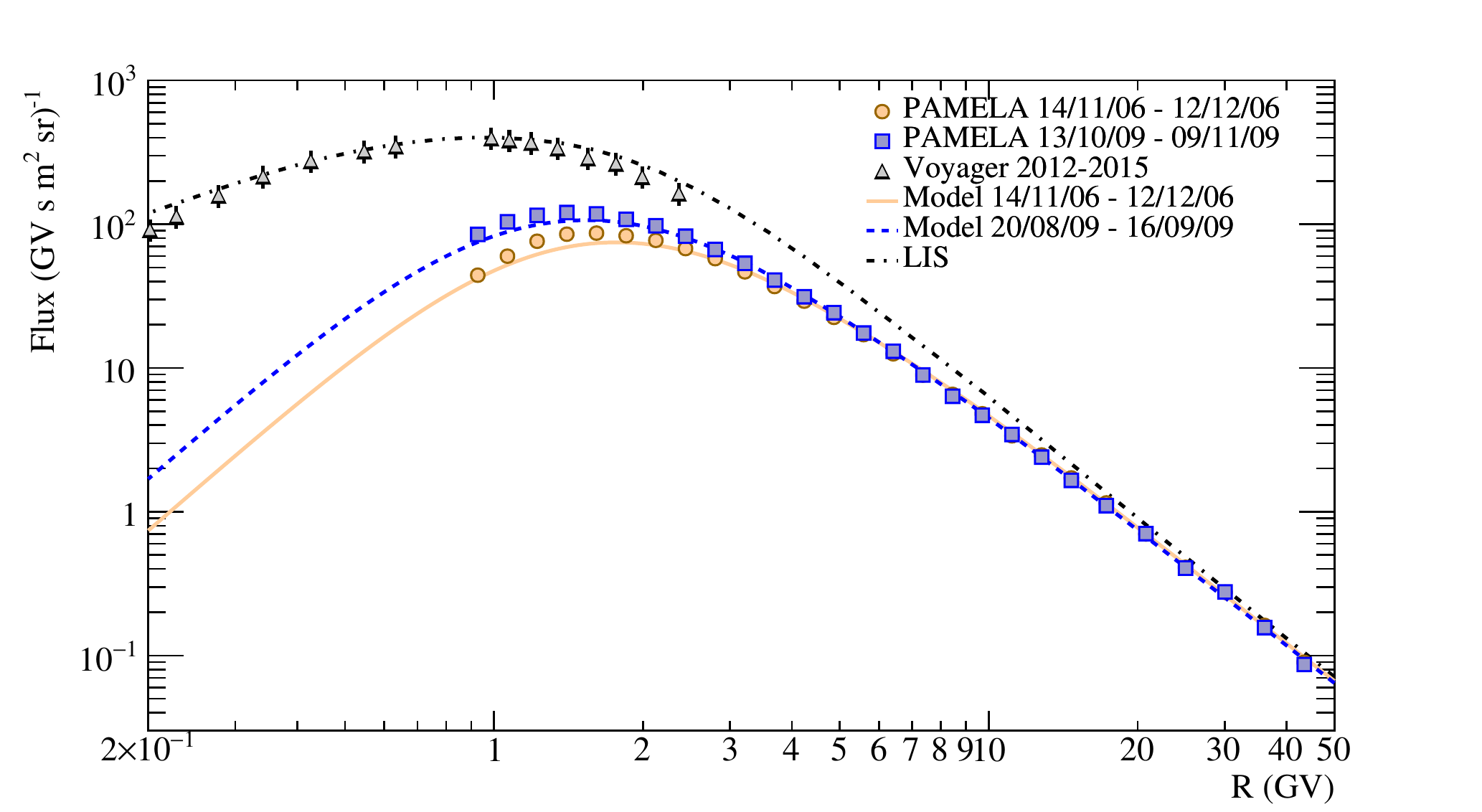}
\caption{Computed helium fluxes as a function of rigidity for November 2006 and October 2009 (solid and dashed lines) compared 
with the measured fluxes by PAMELA (circles and squares). 
The dot-dashed line represents the LIS specified at 120 astronomical units used as input for the model. 
The triangular points are the fluxes measured by Voyager 1 outside the heliosphere \citep{Cummings_2016}.   }
\label{fig_mod}
\end{figure}

\section{\textit{Data Interpretation and Discussion}}

A full three-dimensional cosmic-ray propagation
model for describing CR modulation throughout the heliosphere, already applied to reproduce the PAMELA proton, electron and positron spectra, 
was used to determine the differential intensity of cosmic-ray helium nuclei at Earth from $20$ MV
to $50$ GV. 
This steady-state model is based on the numerical solution of the
Parker transport equation and has been extensively described, and
applied to interpreting observations, by \cite{pot13,pot15,difelice} and recently by \cite{Aslam_2019,Cor19}.

Following the approach of \cite{bish2019} , the LIS computation was performed s
eparately for $^3$He and $^4$He. 
First, $^3$He and $^4$He LIS were obtained with the GALPROP code
\citep{mos98} and then modified 
in order to reproduce the low energy Voyager 1 helium nuclei fluxes measured when the spacecraft entered the local interstellar medium 
in August 2012 \citep{Cummings_2016}.  
The diffusion coefficients used for the computation, as well as all the other 
relevant heliospheric parameters, were essentially the same as obtained for the proton modeling described in \cite{SolarMod}. 
The modulated $^3$He and $^4$He spectra, were converted from kinetic energy per nucleon to rigidity and added together 
to obtain the total helium nuclei spectra presented in Figure  \ref{fig_mod}.  

The solid and the dash lines shown in Figure \ref{fig_mod} represent
the computed spectra for helium nuclei respectively for the Carrington rotation number 2050 (14 November - 12 December 2006) and 2089 (13 October - 9 November 
2009). 
 The model is in agreement with the measurements within the statistical and systematic errors for 2006 and 2009 above 2.5 GV.  Below this energy 
the model is systematically lower by $10-15 \%$ with respect to observational data. It is emphasized that the heliospheric parameters
for helium modulation cannot be changed arbitrarily from those for protons to improve the quantitative agreements between the modelling results
and these observations. There is no compelling reason to believe that GCR helium is modulated in a fundamentally different way with respect from GCR protons.  
However, some adjustments may be made to the LIS for helium, but because it is constrained by the mentioned Voyager 1 observations 
at low rigidity and by PAMELA observations at high rigidity where modulation is negligible, only minor adjustments have been made to the
one published by \cite{bish2019} to obtain a LIS as shown in Figure \ref{fig_mod}. Therefore the LIS is assumed to be optimal in reproducing the observed spectra
for the two periods at Earth over this wide rigidity range. Qualitatively, the modelling reproduces how 
the spectra became gradually softer as more low energy particles arrive at the Earth with decreasing solar activity to reach the maximum value in December 2009.

When the ratio of the computed proton and helium fluxes is calculated as a function of time, it gives an overall decrease from  2006 to 2009 of about 
$20\%$ at $200$ MV while no significant variation is predicted above $\sim 2$ GV. This results are in agreement with the observed ratios.

If there is no fundamental difference between proton and helium modulation in the heliosphere, a non-constant proton-to-helium 
flux ratio over time, for a given rigidity, is the result of: 
\begin{enumerate}
\item their different velocities, related to the different particle masses
(which play a role in the effective rigidity dependence of the various diffusion coefficients as well as the drift coefficient);
\item the different shape of the proton LIS compared to the helium LIS.
\end{enumerate}

 Using the proton and helium LIS from \cite{bish2019} it can be
noticed that the different shapes of the two LIS result in a
proton-to-helium  
flux ratio that, below 30 GV, increases
progressively with decreasing rigidity, even significantly below 2 GV.
When this ratio additionally changes with time, it means that the difference in the slopes of these LISs causes the effects 
of adiabatic energy losses to be somewhat different for helium than for protons when the overall modulation conditions change, as happened from 2006 to 2009. 
Moreover, the flux of CRs with lower rigidity reaches a maximum value at Earth later than the flux with higher rigidity when minimum solar activity 
had occurred. Following maximum solar activity, low rigidity CRs will reach a minimum value first, see e.g \cite{Roux1,Pot3}.

A detailed description of the modelling 
along with a more quantitative study of the relative
contribution of  
1) and 2) to the observed time-dependent proton-to-helium flux
ratio will be presented in a future
publication. 
The analysis of the PAMELA data is still in progress to
extend the time-dependent spectra of the helium nuclei  
 to the $24^{th}$ solar maximum up to the reversal of heliospheric magnetic field polarity. This will cover the gap 
 and overlap with the results recently published by the AMS-02
collaboration that reported a time-dependent proton-to-helium flux ratio between $\approx 2$ and 3 GV in the period between 2011 and 2017 \citep{PhysRevLett.121.051101}.

The results discussed in this paper will be available at the
Cosmic Ray Data Base of the ASI Space Science Data Center
(http://tools.asdc.asi.it/CosmicRays/chargedCosmicRays.jsp).

\acknowledgments

We acknowledge partial financial support from The Italian Space Agency (ASI) under the program ``Programma PAMELA - attivit\`a scientifica di analisi dati in fase E''.
We also acknowledge support from Deutsches Zentrum fur Luft- und Raumfahrt (DLR), The Swedish National Space Board, The Swedish Research Council, The Russian Space Agency (Roscosmos).


\begin{deluxetable}{ccccc}
\tabletypesize{\footnotesize}
\tablecolumns{5}
\tablewidth{0pc}
\tablecaption{Helium nuclei fluxes measured by PAMELA over four time periods (Carrington rotations 2050, 2064, 2077 and 2088-2091). The first and second errors represent the statistical
and systematic errors, respectively. \label{tab1}    }
\tablehead{
\colhead{Kinetic Energy } &  \multicolumn{4}{c}{Flux} \\
\colhead{(GeV/n)} &  \multicolumn{4}{c}{(particles/(m$^{2}$ sr
s GeV))} \\ 
\cline{2-5} \\
\colhead{} & \colhead{2006/11/14-2006/12/12} & 
\colhead{  2007/12/01-2007/12/29 } & \colhead{ 2008/11/20-2008/12/17} &
\colhead{ 2009/09/16-2010/01/03}}
\startdata
0.095-0.124 & (1.99  $\pm$ 0.05  $\pm$ 0.05)$  \times 10^{2}$ &   (2.56  $\pm$ 0.05 $\pm$ 0.06)$ \times 10^{2}$   &   (2.94 $\pm$ 0.08 $\pm$ 0.08)$ \times 10^{2}$ 	   &   (4.08 $\pm$ 0.14 $\pm$ 0.22)$ \times 10^{2}$ \\
0.124-0.160 & (2.41  $\pm$ 0.05  $\pm$ 0.05)$  \times 10^{2}$ &   (3.09  $\pm$ 0.05 $\pm$ 0.08)$ \times 10^{2}$   &   (3.35 $\pm$ 0.08 $\pm$ 0.10)$ \times 10^{2}$  	   &   (4.31 $\pm$ 0.13 $\pm$ 0.23)$ \times 10^{2}$ \\ 
0.160-0.206 & (2.78  $\pm$ 0.04  $\pm$ 0.07)$  \times 10^{2}$ &   (3.38  $\pm$ 0.05 $\pm$ 0.08)$ \times 10^{2}$   &   (3.67 $\pm$ 0.07 $\pm$ 0.11)$ \times 10^{2}$ 	   &   (4.44 $\pm$ 0.11 $\pm$ 0.23)$ \times 10^{2}$ \\ 
0.206-0.264 & (2.84  $\pm$ 0.04  $\pm$ 0.07)$  \times 10^{2}$ &   (3.32  $\pm$ 0.04 $\pm$ 0.08)$ \times 10^{2}$   &   (3.60 $\pm$ 0.06 $\pm$ 0.10)$ \times 10^{2}$  	   &   (4.15 $\pm$ 0.10 $\pm$ 0.22)$ \times 10^{2}$ \\ 
0.264-0.337 & (2.66  $\pm$ 0.03  $\pm$ 0.06)$  \times 10^{2}$ &   (3.04  $\pm$ 0.04 $\pm$ 0.07)$ \times 10^{2}$   &   (3.29 $\pm$ 0.05 $\pm$ 0.09)$ \times 10^{2}$	   &   (3.81 $\pm$ 0.08 $\pm$ 0.20)$ \times 10^{2}$ \\ 
0.337-0.427 & (2.37  $\pm$ 0.02  $\pm$ 0.06)$  \times 10^{2}$ &   (2.71  $\pm$ 0.02 $\pm$ 0.06)$ \times 10^{2}$   &   (2.84 $\pm$ 0.04 $\pm$ 0.08)$ \times 10^{2}$ 	   &   (3.22 $\pm$ 0.05 $\pm$ 0.17)$ \times 10^{2}$ \\ 
0.427-0.537 & (2.06  $\pm$ 0.03  $\pm$ 0.05)$  \times 10^{2}$ &   (2.34  $\pm$ 0.02 $\pm$ 0.05)$ \times 10^{2}$   &   (2.42 $\pm$ 0.03 $\pm$ 0.07)$ \times 10^{2}$ 	   &   (2.69 $\pm$ 0.05 $\pm$ 0.14)$ \times 10^{2}$ \\ 
0.537-0.670 & (1.71  $\pm$ 0.02  $\pm$ 0.04)$  \times 10^{2}$ &   (1.89  $\pm$ 0.02 $\pm$ 0.04)$ \times 10^{2}$   &   (1.97 $\pm$ 0.02 $\pm$ 0.06)$ \times 10^{2}$   	   &   (2.17 $\pm$ 0.04 $\pm$ 0.12)$ \times 10^{2}$ \\ 
0.670-0.831 & (1.39  $\pm$ 0.01  $\pm$ 0.03)$  \times 10^{2}$ &   (1.50  $\pm$ 0.01 $\pm$ 0.04)$ \times 10^{2}$   &   (1.56 $\pm$ 0.02 $\pm$ 0.04)$ \times 10^{2}$ 	   &   (1.74 $\pm$ 0.03 $\pm$ 0.09)$ \times 10^{2}$ \\ 
0.831-1.023 & (1.08  $\pm$ 0.01  $\pm$ 0.02)$  \times 10^{2}$ &   (1.16  $\pm$ 0.01 $\pm$ 0.03)$ \times 10^{2}$   &   (1.20 $\pm$ 0.01 $\pm$ 0.03)$ \times 10^{2}$  	   &   (1.30 $\pm$ 0.02 $\pm$ 0.07)$ \times 10^{2}$ \\ 
1.02-1.25   & (8.27  $\pm$ 0.07  $\pm$ 0.20)$  \times 10^{1}$ &   (8.81  $\pm$ 0.07  $\pm$ 0.21)$  \times 10^{1}$ &   (9.12  $\pm$ 0.11  $\pm$ 0.26)$  \times 10^{1}$  &   (9.71 $\pm$ 0.17  $\pm$ 0.51)$  \times 10^{1}$ \\
1.25-1.52   & (6.38  $\pm$ 0.06  $\pm$ 0.15)$  \times 10^{1}$ &   (6.73  $\pm$ 0.06  $\pm$ 0.16)$  \times 10^{1}$ &   (6.89  $\pm$ 0.09  $\pm$ 0.20)$  \times 10^{1}$  &   (7.17 $\pm$ 0.13  $\pm$ 0.38)$  \times 10^{1}$ \\ 
1.52-1.83   & (4.83  $\pm$ 0.05  $\pm$ 0.11)$  \times 10^{1}$ &   (5.02  $\pm$ 0.05  $\pm$ 0.12)$  \times 10^{1}$ &   (5.05  $\pm$ 0.07  $\pm$ 0.15)$  \times 10^{1}$  &   (5.34 $\pm$ 0.11  $\pm$ 0.29)$  \times 10^{1}$ \\ 
1.83-2.20   & (3.60  $\pm$ 0.04  $\pm$ 0.09)$  \times 10^{1}$ &   (3.71  $\pm$ 0.04  $\pm$ 0.09)$  \times 10^{1}$ &   (3.72  $\pm$ 0.06  $\pm$ 0.11)$  \times 10^{1}$  &   (3.91 $\pm$ 0.08  $\pm$ 0.20)$  \times 10^{1}$ \\ 
2.20-2.62   & (2.62  $\pm$ 0.03  $\pm$ 0.06)$  \times 10^{1}$ &   (2.72  $\pm$ 0.03  $\pm$ 0.06)$  \times 10^{1}$ &   (2.88  $\pm$ 0.05  $\pm$ 0.08)$  \times 10^{1}$  &   (2.83 $\pm$ 0.07  $\pm$ 0.15)$  \times 10^{1}$ \\ 
2.62-3.11   & (1.84  $\pm$ 0.02  $\pm$ 0.04)$  \times 10^{1}$ &   (1.90  $\pm$ 0.02  $\pm$ 0.04)$  \times 10^{1}$ &   (2.04  $\pm$ 0.04  $\pm$ 0.06)$  \times 10^{1}$  &   (1.97 $\pm$ 0.05  $\pm$ 0.10)$  \times 10^{1}$ \\ 
3.11-3.68   & (1.34  $\pm$ 0.02  $\pm$ 0.03)$  \times 10^{1}$ &   (1.39  $\pm$ 0.02  $\pm$ 0.03)$  \times 10^{1}$ &   (1.47  $\pm$ 0.03  $\pm$ 0.04)$  \times 10^{1}$  &   (1.45 $\pm$ 0.04  $\pm$ 0.07)$  \times 10^{1}$ \\ 
3.68-4.34   &  9.80  $\pm$ 0.15  $\pm$ 0.23 		      &    9.77  $\pm$ 0.15  $\pm$ 0.23 		     &    9.91  $\pm$ 0.21  $\pm$ 0.28  		   &   (1.04 $\pm$ 0.03  $\pm$ 0.05)$  \times 10^{1}$ \\ 
4.34-5.10   &  6.85  $\pm$ 0.09  $\pm$ 0.16 		      &    7.00  $\pm$ 0.10  $\pm$ 0.17  		     &    6.88  $\pm$ 0.14  $\pm$ 0.22 		   &    7.10 $\pm$ 0.21  $\pm$ 0.39 \\ 
5.10-5.97   &  5.02  $\pm$ 0.07  $\pm$ 0.12 		      &    5.13  $\pm$ 0.07  $\pm$ 0.12  		     &    5.27  $\pm$ 0.11  $\pm$ 0.15 		   &    4.94 $\pm$ 0.16  $\pm$ 0.27 \\ 
5.97-6.98   &  3.47  $\pm$ 0.06  $\pm$ 0.08 		      &    3.40  $\pm$ 0.06  $\pm$ 0.08  		     &    3.70  $\pm$ 0.08  $\pm$ 0.10 		   &    3.79 $\pm$ 0.12  $\pm$ 0.19 \\ 
6.98-8.57   &  2.32  $\pm$ 0.04  $\pm$ 0.06 		      &    2.33  $\pm$ 0.04  $\pm$ 0.05  		     &    2.26  $\pm$ 0.06  $\pm$ 0.07 		   &    2.35 $\pm$ 0.08  $\pm$ 0.12 \\ 
8.57-10.46  & 1.41   $\pm$ 0.03  $\pm$ 0.03 		      &    1.41  $\pm$ 0.03  $\pm$ 0.03  		     &    1.41  $\pm$ 0.04  $\pm$ 0.04 		   &    1.41 $\pm$ 0.06  $\pm$ 0.08 \\ 
10.46-12.75 & (8.25  $\pm$ 0.15  $\pm$ 0.20)$ \times 10^{-1}$ &   (8.28  $\pm$ 0.16  $\pm$ 0.21)$ \times 10^{-1}$ &   (8.22  $\pm$ 0.23  $\pm$ 0.25)$ \times 10^{-1}$  &   (8.10 $\pm$ 0.34  $\pm$ 0.46)$ \times 10^{-1}$ \\ 
12.75-15.50 & (5.53  $\pm$ 0.11  $\pm$ 0.13)$ \times 10^{-1}$ &   (5.32  $\pm$ 0.11  $\pm$ 0.13)$ \times 10^{-1}$ &   (5.49  $\pm$ 0.17  $\pm$ 0.17)$ \times 10^{-1}$  &   (5.66 $\pm$ 0.25  $\pm$ 0.32)$ \times 10^{-1}$ \\ 
15.50-18.82 & (3.25  $\pm$ 0.08  $\pm$ 0.08)$ \times 10^{-1}$ &   (3.17  $\pm$ 0.08  $\pm$ 0.08)$ \times 10^{-1}$ &   (2.96  $\pm$ 0.12  $\pm$ 0.10)$ \times 10^{-1}$  &   (3.22 $\pm$ 0.17  $\pm$ 0.18)$ \times 10^{-1}$ \\ 
18.82-22.80 & (1.80  $\pm$ 0.05  $\pm$ 0.04)$ \times 10^{-1}$ &   (1.83  $\pm$ 0.06  $\pm$ 0.05)$ \times 10^{-1}$ &   (2.09  $\pm$ 0.09  $\pm$ 0.07)$ \times 10^{-1}$ &   (1.64 $\pm$ 0.12  $\pm$ 0.10)$ \times 10^{-1}$ 	\\  
\enddata
\end{deluxetable}

\bibliography{Bibtex_solmod}

\end{document}